# Percentile-based probabilistic optimization: a feasibility study in IMPT head-and-neck cancer patients

Jenneke I. de Jong [a,b], Sebastiaan Breedveld [a], Albin Fredriksson [c], Johan Sundström [c], Zoltan Perko [d], Danny Lathouwers [d], Jelte R. de Jong [d], Erik Engwall [c], Steven J.M. Habraken [e,b], Mischa S. Hoogeman [a,b].

a. Radiotherapy, Erasmus MC Cancer Institute, Rotterdam, the Netherlands.
b. Medical Physics & Informatics, HollandPTC, Delft, the Netherlands.
c. R&D department, RaySearch Laboratories, Stockholm, Sweden
d. Radiation Science and Technology, Delft University of Technology, Delft, the Netherlands.
e. Radiation Oncology, Leiden University Medical Center, Leiden, the Netherlands.

---

**Abstract**

*Objective*

We evaluated whether direct percentile-based probabilistic optimization can steer intensity-modulated proton therapy (IMPT) plans toward clinical goals at specified probability levels and improve the trade-off between target coverage and organ-at-risk (OAR) sparing compared with minimax optimization.

*Approach*

Eight oropharyngeal and four nasopharyngeal cancer patients were included. Direct probabilistic optimization accounted for systematic and random setup and systematic range uncertainties by simulating 1,000 fractionated treatment courses per optimization iteration and directly optimizing percentile-based dose-volume metrics. Optimization targeted 10th-percentile $D_{99.8\%,CTV}$=66.50 Gy RBE (oropharyngeal) or a critical OAR constraint (nasopharyngeal). Minimax optimization based on 21 scenarios calibrated per patient using probabilistic evaluation to steer toward the same probabilistic goals. Experiment 1 characterized target coverage–critical-OAR sparing trade-offs using simplified plans. Experiment 2 compared clinically representative plans at matched probabilistic target coverage (oropharyngeal) or critical-OAR dose (nasopharyngeal).

*Main results*

In experiment 1, direct probabilistic optimization improved the trade-off between target coverage and mandible sparing in seven of eight oropharyngeal patients compared with probabilistically calibrated minimax optimization. It did not improve the trade-off between target coverage and critical-OAR dose in any of the three nasopharyngeal patients. In experiment 2, direct probabilistic optimization steered plans close to predefined probabilistic goals, with a median 10th-percentile $D_{99.8\%,\text{primary CTV}}$ of 66.60 Gy RBE (66.56–66.77). At matched target coverage, OAR dose decreased in seven of eight oropharyngeal plans, with median reductions of 0.70 Gy RBE for 95th-percentile $D_{2\%,\text{mandible}}$ and 0.15 Gy RBE in mean non-critical-OAR dose. In the nasopharyngeal patients, changes in 10th-percentile $V_{0.95\text{Dpres},\text{primary CTV}}$ at matched critical-OAR dose were small and inconsistent, ranging from −0.79 to +0.85 percentage points.

*Significance*

Direct probabilistic optimization can steer IMPT plans toward specified probabilistic goals and improve OAR sparing at comparable target coverage in oropharyngeal patients. No consistent dosimetric benefit was demonstrated in nasopharyngeal patients, which may indicate a need for further refinement when optimizing extreme dose-volume metrics for small OARs.

## Introduction

Intensity-modulated proton therapy (IMPT) is very sensitive to treatment uncertainties, including setup and range errors [1-3]. In photon therapy, uncertainties are commonly addressed by expanding the clinical target volume (CTV) to the planning target volume (PTV) with a margin [4]. For IMPT, this approach is inadequate because uncertainties can result in changes in the shape of the dose distribution due to the dependence of the proton range on tissue composition and the steep dose gradients [5]. Therefore, robust optimization and robustness evaluation are required to ensure adequate target coverage while limiting dose to organs at risk (OARs) [6].

Different optimization strategies have been proposed to generate robust IMPT plans, often based on a set of predefined error scenarios [7]. Different strategies have been proposed for incorporating these scenarios into the optimization process, including worst-case and scenario-based margins [8-10]. A widely used approach is scenario-wise minimax optimization [11], which optimizes the treatment plan with respect to the worst-performing error scenario within the predefined error scenario set.

In clinical practice in the Netherlands, robust optimization is performed using the minimax method based on a predefined set of 21 error scenarios. These consist of the nominal scenario and positive and negative setup errors along the three principal directions, combined with range errors of −3%, 0%, and +3%. Robustness evaluation is subsequently performed using 28 error scenarios, consisting of 14 setup error scenarios distributed over the surface of a sphere, each combined with range errors of -3% and +3%. Target coverage is then assessed using the voxel-wise minimum (VWmin) dose distribution, constructed by taking the minimum dose in each voxel across all 28 dose distributions [12]. This approach ensures robustness against the chosen uncertainty scenarios but does not explicitly quantify clinical goal values at specified probability levels.

Previous work has combined the clinically used minimax optimization method with probabilistic evaluation, which explicitly incorporates the probability distributions of random setup, systematic setup and range errors [13]. Compared with planning based on VWmin-goals, this approach can improve the trade-off between OAR sparing and target coverage [13] and reduce interpatient variation in probabilistic outcomes [14]. However, minimax optimization uses a discrete worst-case criterion, whereas probabilistic evaluation considers clinical goal outcomes at specified probability levels, leaving the optimization and evaluation criteria misaligned. As a result, the minimax robustness setting requires patient-specific calibration to meet the clinical goals at a desired probability level. In this study, we refer to this approach as probabilistically calibrated minimax optimization.

Probabilistic optimization can address this mismatch by incorporating the probability distributions of treatment uncertainties directly into the optimization process [15]. Proposed approaches include optimization of the expected objective value [5], expected dose and dose variance [16], probability-weighted discrete scenarios [17], probabilistic dose-coverage criteria [18, 19] and direct optimization of voxel-wise dose percentiles corresponding to specified probabilities of target underdosage or OAR overdosage [20, 21].

However, important barriers to clinical implementation remain. Probabilistic optimization is computationally demanding due to the large number of dose distributions required. In addition, fractionation and random errors are often not explicitly accounted for during optimization, potentially limiting the accurate simulation of treatment courses. Furthermore, not all methods directly optimize clinical goals at specified probability levels, and most have only been evaluated in methodological studies or small patient cohorts.

Recently, Fredriksson et al. developed a percentile-based probabilistic optimization method that explicitly accounts for systematic and random uncertainties and simulates fractionated treatment courses [15]. The method is computationally efficient by sampling from a grid of precalculated dose distributions, an approach previously validated for probabilistic evaluation in [22]. However, this approach is limited to the setup and range errors represented in the grid, and whether these sufficiently capture the relevant uncertainties remains to be investigated. The method was previously demonstrated successfully in a pencil beam scanning proton brain case [15], but its ability to consistently steer treatment plans toward specified probabilistic clinical goals across patients and under clinically representative planning conditions remains to be investigated.

The aim of this study is to assess whether this direct probabilistic optimization method can consistently steer treatment plans toward meeting clinical goals at specified probability levels within a cohort of head and neck cancer patients. In addition, we evaluate whether direct probabilistic optimization provides a more favorable trade-off between OAR sparing and target coverage than probabilistically calibrated minimax optimization.

---

## Methods

### Patient cohort and study design

Twelve head and neck cancer (HNC) patients treated at HollandPTC in 2024 were included in this study, including eight patients with oropharyngeal cancer and four with nasopharyngeal cancer.

The prescribed dose ($D_{pres}$) was 70 Gy RBE for the primary CTV and 54.25 Gy RBE to the elective lymph nodes, delivered in 35 fractions. All plans were created in a research version of RayStation v2025 using the clinical beam model and the Monte Carlo (MC) dose engine with a MC uncertainty of 1%. Dose was reported in Gy RBE assuming a constant RBE of 1.1. Intensity-modulated proton therapy (IMPT) plans were generated using six beams.
For each patient, plans were generated using two optimization approaches: probabilistically calibrated minimax optimization and direct probabilistic optimization. The two optimization approaches were compared in two experiments. In Experiment 1, simplified treatment plans were generated to characterize the trade-off between probabilistic CTV coverage and sparing of a single patient-specific critical OAR. To isolate this trade-off, objectives and constraints for the remaining OARs were excluded. In Experiment 2, clinically representative treatment plans were generated by including the additional OAR objectives used in clinical practice, allowing the optimization approaches to be compared under more realistic clinical planning conditions.

**Uncertainty distributions**

Systematic and random setup errors were assumed to follow zero-mean Gaussian distributions with standard deviations $\Sigma$ = 0.92 mm and $\sigma$ = 1.00 mm, respectively. Range errors were modeled as systematic errors following a zero-mean Gaussian distribution with a standard deviation of 1.5% of the nominal stopping power. Both systematic setup errors and range errors were assumed to remain constant across all fractions of a treatment course, whereas random setup errors were modeled as independent errors sampled separately for each fraction. In the dose calculations, range uncertainties were simulated by scaling the patient mass density.

The values of $\Sigma$ and $\sigma$ were chosen to be consistent with the $M$ = 3.00 mm margin used at our center according to the van Herk recipe ($M = 2.5\Sigma + 0.7\sigma$) [4] and are close to values obtained in previous measurements at our center [23]. The standard deviation of the range error was selected to be consistent with values reported in the literature [24]. Gaussian distributions were adopted as a modeling assumption, consistent with commonly used uncertainty models in radiotherapy.

**Sparse probabilistic evaluation**

During and after optimization, probabilistic evaluation was performed using the sparse probabilistic evaluation (SPE) method (Figure 1A) [22], which enables efficient estimation of DVH metric values at specified probability levels. SPE relies on a predefined setup- and range-error grid with corresponding dose distributions. During optimization, these dose distributions were calculated using precalculated dose-influence data combined with the updated spot weights to reduce computational time. For probabilistic evaluation after optimization, the dose distributions were calculated using the full MC dose calculation method.

In this study, an *error scenario* refers to a single fraction dose distribution corresponding to a combination of setup and range errors for a given set of beamlet weights. In contrast, a *simulated treatment course* refers to the accumulated dose distribution over all 35 fractions, obtained by sampling treatment errors, selecting the corresponding error-scenario dose distribution for each fraction, and summing the resulting fraction doses.

Assuming independent systematic and random setup errors, the standard deviation of the combined setup error was calculated as SD = $\sqrt{\Sigma^2 + \sigma^2}$. For the setup-error grid, five evenly spaced points between −3 SD and +3 SD were selected along each principal direction, resulting in 125 grid points in 3D. The grid was subsequently truncated to a sphere with a radius of 3 SD, leaving 33 setup-error points. These points were combined with seven range-error values spanning −4.5% to +4.5% in increments of 1.5%, resulting in 231 dose distributions.

We simulated 1,000 treatment courses of 35 fractions per patient. For each simulated treatment course, one systematic setup error and one range error were sampled from their respective Gaussian distributions and kept constant across all fractions. Additionally, a random setup error was sampled independently for each fraction. For each fraction, the sum of the random and the systematic setup error was mapped to the nearest error point in the predefined setup error grid. The sampled range error was mapped to the nearest point in the range error grid. The dose distribution corresponding to both the selected setup error point and range error point was then assigned to that fraction. Dose distributions were accumulated over all 35 fractions to obtain the dose distribution for each simulated treatment course.

For the oropharyngeal cancer patients, the accuracy of SPE in this patient cohort was previously validated against full MC dose calculations performed for each fraction, without relying on a predefined error grid [22]. As nasopharyngeal cancer patients were not included in the previous validation study, SPE was further evaluated for a subset of nasopharyngeal treatment plans by comparison with estimated probabilistic outcomes obtained using polynomial chaos expansion (PCE) [Supplementary Materials 1]. PCE was used only for this additional validation and not during optimization or the primary plan comparisons. PCE is an established method for probabilistic evaluation that has been widely applied in research and was previously validated for radiotherapy application [25]. SPE showed close agreement with PCE, with absolute differences of ≤0.28 Gy RBE (<0.4% of Dpres) for the evaluated dose metrics and ≤0.18 percentage points for the 10th-percentile $V_{0.95\text{Dpres,primary CTV}}$ [Supplementary Materials 1].

**Direct probabilistic optimization**

The direct probabilistic optimization method (Figure 1A) was based on the method described in [15]. During direct probabilistic optimization, SPE was performed in every iteration, simulating 1,000 treatment courses. All SPE settings were identical to those described in the *sparse probabilistic evaluation* section, except that a different random seed was used for uncertainty sampling. For each probabilistic objective or constraint separately, the simulated treatment courses were ranked according to the corresponding objective-function/constraint value**.** For a probabilistic objective or constraint with a specified probability level $\alpha$, the worst-performing 100-$\alpha$% of treatment courses were discarded, and optimization proceeded using the remaining $\alpha$% of treatment courses. Following this selection, the objective-function values of the retained $\alpha$% of treatment courses were aggregated for each objective or constraint using a generalized mean. For the probabilistic constraints, a power of 1 (arithmetic mean) was used. Because the constraint-function values are nonnegative and equal to zero when the constraint is satisfied, an arithmetic mean of zero ensures that all retained treatment courses satisfy the constraint. For probabilistic objectives, a power of 8 was used, corresponding to the value employed in RayStation's minimax optimization. Further details on the aggregation procedure are provided in Supplementary Materials 7.

Thus, for a probability level of 90%, the worst-performing 10% of treatment courses were excluded from the generalized mean, and the optimization round was continued using the remaining 90%. Because the ranking was performed separately for each objective or constraint, the subset of treatment courses included in the optimization could differ between objectives and constraints. This process was repeated at every iteration until convergence of the optimization was reached.

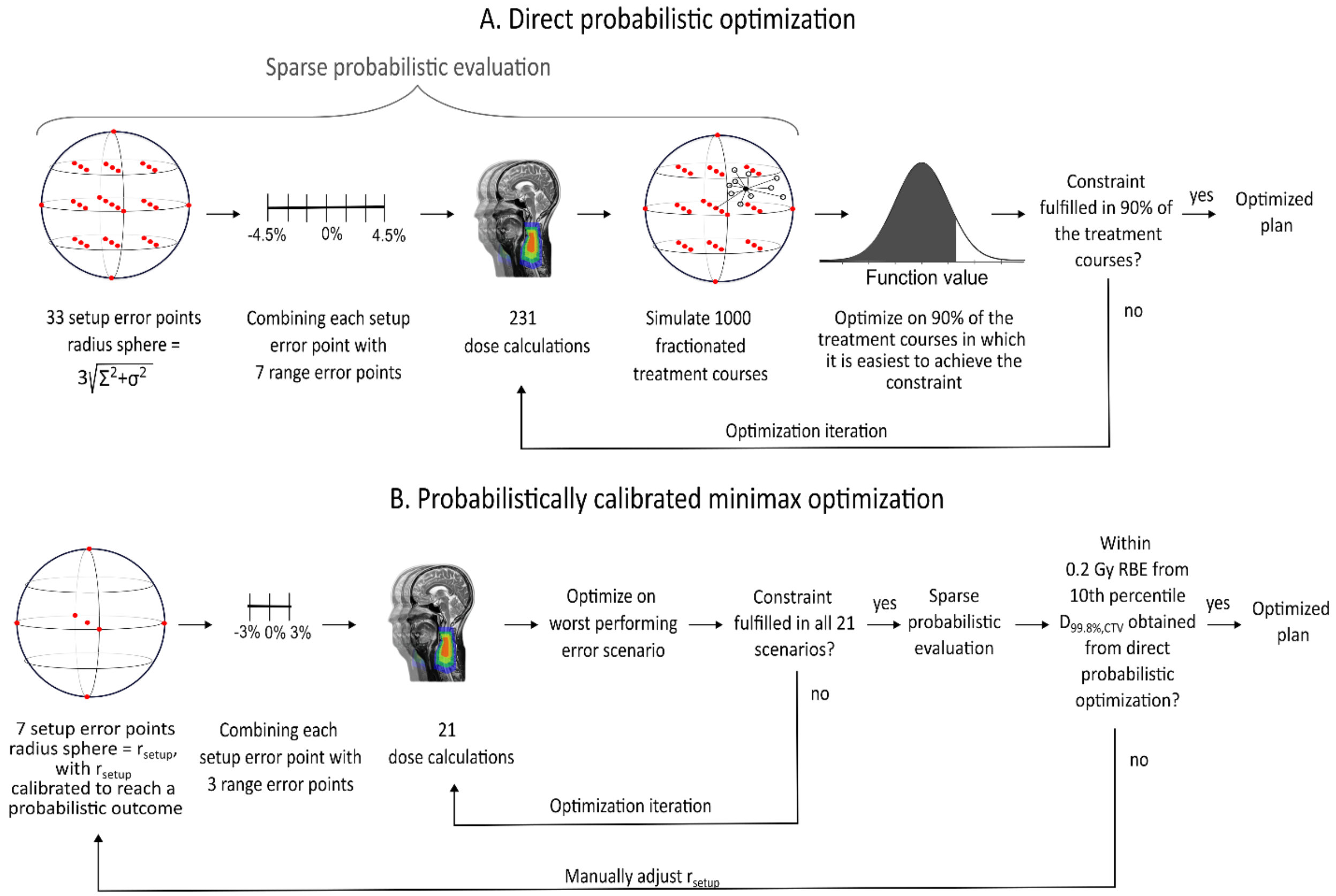

***Figure 1.*** *Schematic overview of the two optimization approaches for the oropharyngeal cancer patients. The calibration procedure can also be applied to other probabilistic endpoints, such as critical-OAR dose metrics. (A)* ***Direct probabilistic optimization.*** *Sparse probabilistic evaluation (SPE) uses a predefined grid of setup and range error scenarios with Monte Carlo-calculated dose distributions to simulate 1,000 fractionated treatment courses. For each probabilistic objective or constraint, the simulated treatment courses are ranked according to the corresponding objective-function value, and optimization is performed using the fraction of treatment courses defined by the specified probability level. In the illustrated example, a probability level of 90% is used, such that the worst-performing 10% of treatment courses are excluded and optimization is based on the remaining 90%. This procedure is repeated during optimization. (B****) Probabilistically calibrated minimax optimization****. Minimax optimization is performed using 21 predefined error scenarios, with the*

*illustrated constraint required to be satisfied in all scenarios. Sparse probabilistic evaluation is subsequently performed, and the setup robustness parameter $r_{setup}$ is manually adjusted and the plan reoptimized until the predefined probabilistic criterion is met. This outer calibration loop was introduced for this study and is not part of the standard clinical minimax optimization procedure.*

**Probabilistically calibrated minimax optimization**

Probabilistically calibrated minimax optimization (Figure 1B) was used as the comparator to direct probabilistic optimization. In this approach, conventional minimax optimization was calibrated using probabilistic evaluation to achieve a selected probabilistic endpoint.

Minimax optimization was based on a set of 21 error scenarios. The scenario set comprised seven setup scenarios, consisting of the nominal setup and shifts along each of the three principal directions defined by the plan-specific setup robustness parameter $r_{setup}$, combined with three range-error scenarios (−3%, 0% and +3%). Because minimax optimization does not explicitly optimize a probabilistic treatment outcome, $r_{setup}$ was manually tuned for each plan using probabilistic evaluation until the selected probabilistic endpoint matched that obtained with direct probabilistic optimization. Starting from $r_{setup}$ = 2.0 mm, the setup robustness parameter was adjusted upward or downward in increments of 0.1 mm based on the probabilistic evaluation result.

For the oropharyngeal cancer plans, the value of $r_{setup}$ was selected such that the 10th-percentile $D_{99.8\%,CTV}$ differed by less than 0.2 Gy RBE from the value obtained with direct probabilistic optimization. For the nasopharyngeal cancer plans, the value of $r_{setup}$ was selected such that the 95th-percentile $D_{0.03cc,critical\ OAR}$ differed by less than 0.4 Gy RBE from the value obtained with direct probabilistic optimization. For readability, *probabilistically calibrated minimax optimization* is referred to as *minimax optimization* in the Results section.

**Experiment 1: Simplified trade-off analysis**

To further investigate differences between the optimization approaches, simplified treatment plans were generated to characterize the trade-off between probabilistic CTV coverage and sparing of a single patient-specific critical OAR. To isolate this trade-off, all OAR-related objectives and constraints other than those for the patient-specific critical OAR were removed. An example of the complete set of optimization objectives and constraints for one oropharyngeal patient and one nasopharyngeal patient is provided in Supplementary Materials 2. The list is the same for probabilistically calibrated minimax optimization and direct probabilistic optimization.

*Oropharyngeal cancer plans*

For the eight oropharyngeal cancer patients, the mandible was selected as the critical OAR. The simplified optimization problem included two target-coverage constraints: $D_{99.8\%,primary\ CTV} \geq 67$ Gy RBE and $D_{99.8\%,nodal\ CTV} \geq 52$ Gy RBE, each at a probability level of 90%. For the mandible, the initial objective was $D_{2\%,mandible} \leq 60$ Gy RBE at a probability level of 95%.

An initial plan was generated with each optimization approach using these objectives and constraints. To characterize the trade-off between target coverage and mandible sparing, the dose level of the mandible objective was subsequently reduced in 5-Gy RBE increments. Although this progressively sets the objective to dose levels below what is feasible, it increases the optimization pressure on mandible sparing. For each dose level and optimization approach, optimization was started from the corresponding initially optimized plan. This procedure was continued until 10th-percentile $D_{99.8\%,\text{primary CTV}}$=66.5 Gy RBE could no longer be satisfied. The resulting series of plans was used to characterize the trade-off between probabilistic target coverage and mandible sparing.

*Nasopharyngeal cancer plans*

For each of the four nasopharyngeal cancer patients, the most critical OAR was identified based on the dose metric that was closest to its corresponding clinical constraint in the clinical plan. The considered OAR constraints were: $D_{\text{max,brainstem surface}}$, $D_{\text{max,brainstem core}}$, $D_{\text{max,optic chiasm}}$, $D_{\text{max,optic nerve right}}$ or $D_{\text{max,optic nerve left}}$, with clinical constraints of 60, 54, 55, 55 and 55 Gy RBE, respectively. For the patient-specific critical OAR, the corresponding constraint was specified at a probability level of 95%. The two target-coverage objectives were $D_{99.8\%,\text{primary CTV}}\geq 67$ Gy RBE and $D_{99.8\%,\text{nodal CTV}}\geq 52$ Gy RBE, both at a probability level of 90%.

For the patient-specific critical OAR, the optimization constraint was initially set to its corresponding clinical constraint. The constraint was subsequently relaxed in 2-Gy RBE increments, thereby progressively reducing the optimization pressure on OAR sparing and allowing greater optimization freedom for target coverage. A total of five trade-off steps were evaluated for each patient. Although OAR doses above the clinical constraint would not be clinically acceptable, these relaxed constraint levels were included solely to characterize the trade-off between probabilistic target coverage and OAR sparing. Evaluating multiple points along the trade-off curve allowed systematic differences between the two optimization approaches to be assessed rather than relying on a comparison at a single point.

Each successive dose level, corresponding to a 5 Gy RBE decrease in the mandible objective for oropharyngeal plans or a 2 Gy RBE increase in the critical OAR constraint for nasopharyngeal plans, was defined as a trade-off step.

*Plan comparisons*

For each patient and optimization approach, the resulting series of plans was used to construct a trade-off curve. For the oropharyngeal cancer plans, the 10th-percentile $D_{99.8\%,\text{primary CTV}}$ was plotted against the 95th-percentile $D_{2\%,\text{mandible}}$ (Figure 2A). For the nasopharyngeal cancer plans, the 10th-percentile $V_{0.95\text{Dpres, CTV}}$ was plotted against the 95th-percentile $D_{0.03\text{cc,critical OAR}}$ (Figure 2B). $V_{0.95\text{Dpres, CTV}}$ was selected because the predefined target-coverage goal could not be achieved, making the volume receiving at least 95% of the prescribed dose more informative than the near-minimum target dose.

The primary comparison between the two optimization approaches was based on plans generated at the same trade-off step. For both optimization approaches, primary-CTV dose inhomogeneity was also evaluated as a supporting measure using the nominal $D_{2\%,\text{primary CTV}}$.

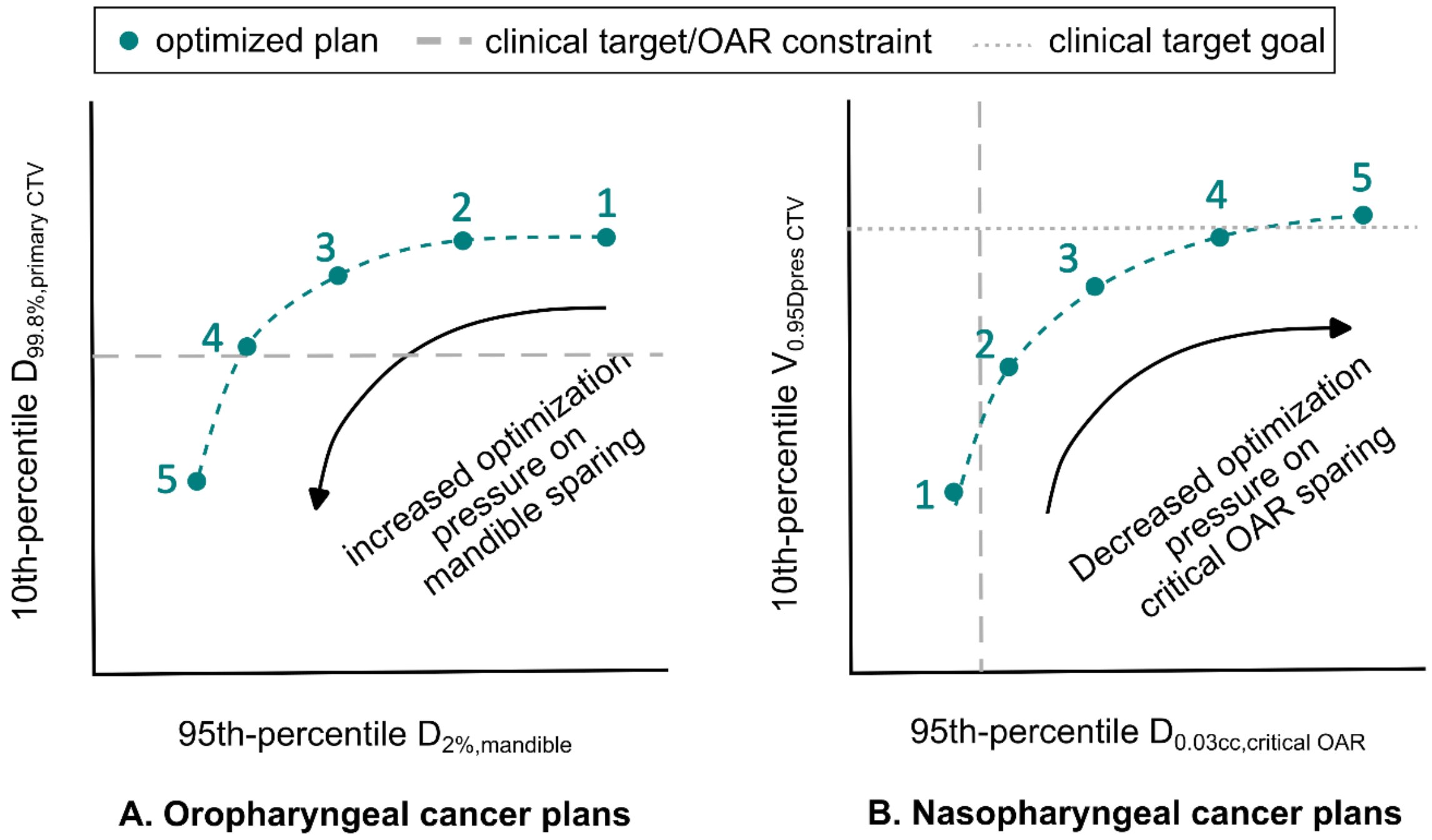

***Figure 2.*** *Schematic illustration of the simplified trade-off analysis used in Experiment 1. Simplified optimization problems were generated by retaining the probabilistic target-coverage constraints and objectives for a single patient-specific critical OAR. Each numbered point represents a successive trade-off step along the curve. (A) Oropharyngeal cancer plans: the mandible objective was progressively strengthened by reducing its dose level, thereby increasing mandible sparing until the predefined probabilistic target-coverage goal was no longer satisfied. (B) Nasopharyngeal cancer plans: the constraint for the patient-specific critical OAR was progressively relaxed, thereby reducing the optimization pressure on OAR sparing and allowing increased probabilistic target coverage.*

**Experiment 2: clinically representative plan comparison**

To compare the two optimization approaches under clinically representative planning conditions, treatment plans were generated using the relevant clinical target and OAR objectives. Mean-dose objectives for the non-critical OARs were included, using the same dose values as specified in the clinical treatment plans. An example of the complete set of optimization objectives and constraints for one oropharyngeal patient and one nasopharyngeal patient is provided in Supplementary Materials 2. The list is the same for probabilistically calibrated minimax optimization and direct probabilistic optimization.

*Oropharyngeal cancer plans*

For each patient, a single mandible objective dose level was selected based on the results of the trade-off analysis performed in Experiment 1. The mandible objective was $D_{2\%,mandible} \leq k$ Gy RBE at a probability level of 95%, with patient-specific values of k of 35, 50, 55 or 60 Gy RBE. For four of the eight patients, an alternative objective configuration was also explored. In this configuration, an upper constraint of $D_{2\%,mandible} \leq 70$ Gy RBE was applied, and the remaining optimization freedom was used to minimize the mean doses to the non-critical OARs.

*Nasopharyngeal cancer plans*

For the four nasopharyngeal cancer patients, six OAR constraints were included: $D_{max,brainstem\ surface} \leq 60$ Gy RBE, $D_{max,brainstem\ core} \leq 54$ Gy RBE, $D_{max,optic\ chiasm} \leq 55$ Gy RBE, $D_{max,optic\ nerve\ left} \leq 55$ Gy RBE, $D_{max,optic\ nerve\ right} \leq 55$ Gy RBE and $D_{2\%,mandible} \leq 70$ Gy RBE. Each constraint was required to be satisfied with a probability level of 95%.

The two target-coverage objectives were $D_{99.8\%,primary\ CTV} \geq 67$ Gy RBE and $D_{99.8\%,nodal\ CTV} \geq 52$ Gy RBE, each at a probability level of 90%. Objectives for CTV conformity and homogeneity were retained from the clinical plan.

*Plan comparisons*

To enable comparison at exactly matched probabilistic endpoints, the dose distributions from both optimization approaches were scaled to remove small residual differences in the matched endpoint. The oropharyngeal cancer plans were matched at a 10th-percentile $D_{99.8\%,primary\ CTV}$ of 66.5 Gy RBE, which also corresponds to 10th-percentile $V_{0.95Dpres}$=99.8%. For the nasopharyngeal cancer plans, plans were scaled so that the 95th-percentile $D_{0.03cc,critical\ OAR}$ was equal to the corresponding clinical constraint value.

The main comparison after matching was the 95th-percentile $D_{2\%,mandible}$ for the oropharyngeal cancer plans and the 10th-percentile $V_{0.95Dpres,primary\ CTV}$ for the nasopharyngeal cancer plans. Additional descriptive comparisons included the nominal mean doses to the remaining OARs, the nominal $D_{2\%,primary\ CTV}$ as a measure of target-dose inhomogeneity and the nominal mean dose to the external body contour.

# Results

## Experiment 1: Simplified trade-off analysis

*Oropharyngeal cancer plans*

For the eight oropharyngeal cases, the initial 10th-percentile $D_{99.8\%,primary\ CTV}$ was comparable between direct probabilistic optimization and minimax optimization, with median values of 66.84 and 66.85 Gy RBE, respectively. The calibrated setup robustness parameter $r_{setup}$ had a median value of 1.8 mm (interpatient range: 1.5–2.0 mm). Patient-specific values are provided in Supplementary Materials 3.

An example of the trade-off between probabilistic target coverage and mandible sparing is shown in Figure 3. For this patient, plans generated at the same trade-off step showed lower 95th-percentile $D_{2\%,mandible}$ with direct probabilistic optimization while maintaining comparable probabilistic target coverage and target-dose homogeneity. Although comparison of minimax plan 1 with direct probabilistic plan 3 at approximately equal target coverage suggests a larger improvement in mandible sparing, the latter plan also showed greater primary-CTV dose inhomogeneity ($D_{2\%,\text{primary CTV}}$).

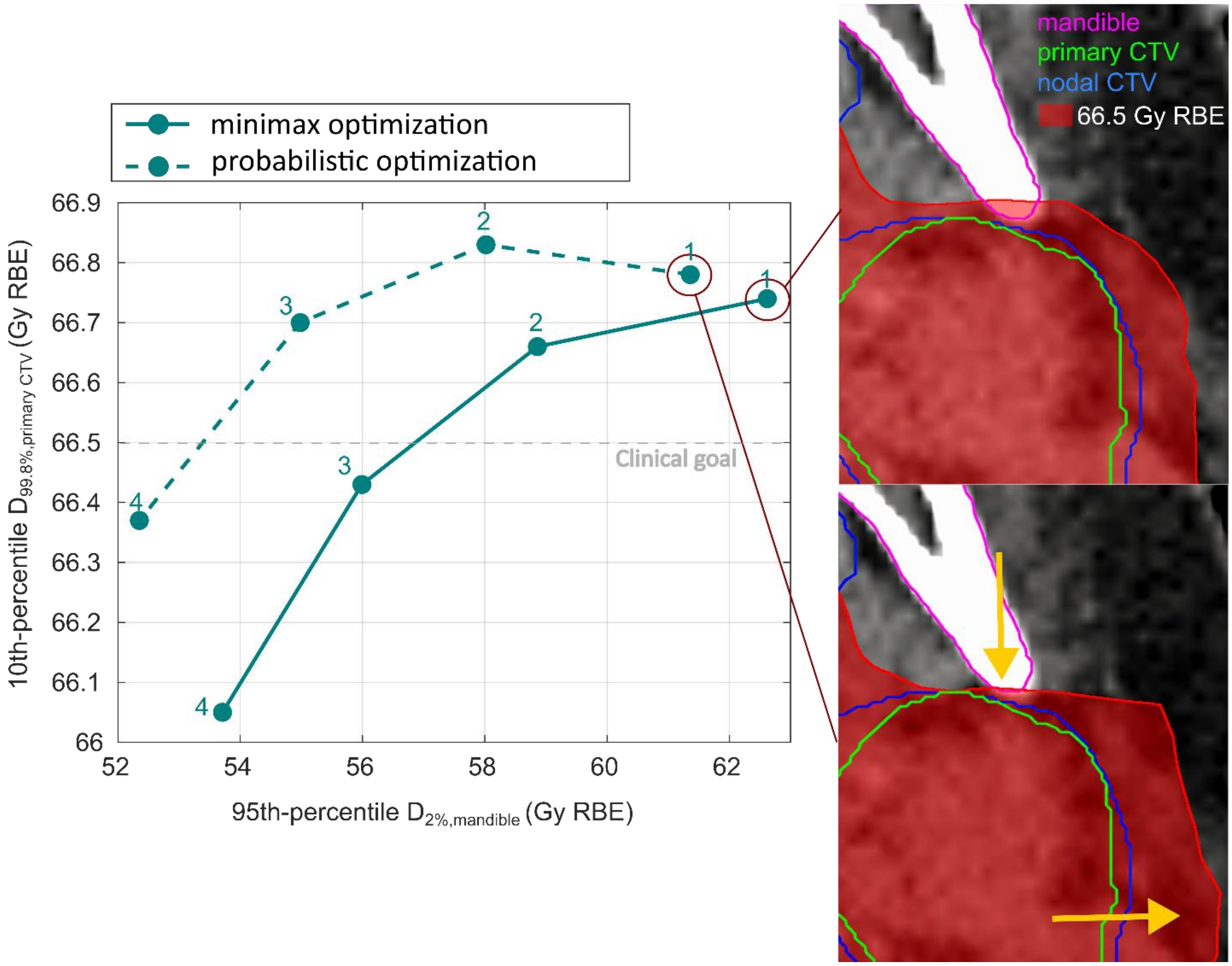

***Figure 3.*** *Example patient for exploring the trade-off between mandible sparing and probabilistic target coverage in simplified oropharyngeal cancer plans. Each numbered point represents a plan generated during the trade-off analysis. Plans labelled with the same number were used for the primary comparison because they were generated at the same trade-off step and had comparable primary-CTV dose homogeneity. Compared with minimax optimization, direct probabilistic optimization produced a dose distribution that conformed more tightly to the CTV near the mandible while extending more broadly in regions where this did not increase mandible dose, as indicated by the yellow arrows.*

The patient-specific trade-off curves for all eight oropharyngeal cancer patients are shown in Figure 4. Plans from the same trade-off step had comparable nominal $D_{2\%,\text{primary CTV}}$ [Supplementary Materials 4]. In the primary comparison between plans from the same trade-off step, direct probabilistic optimization provided a more favorable trade-off between probabilistic target coverage and mandible sparing in seven of the eight patients, with patient 5 being the only exception.

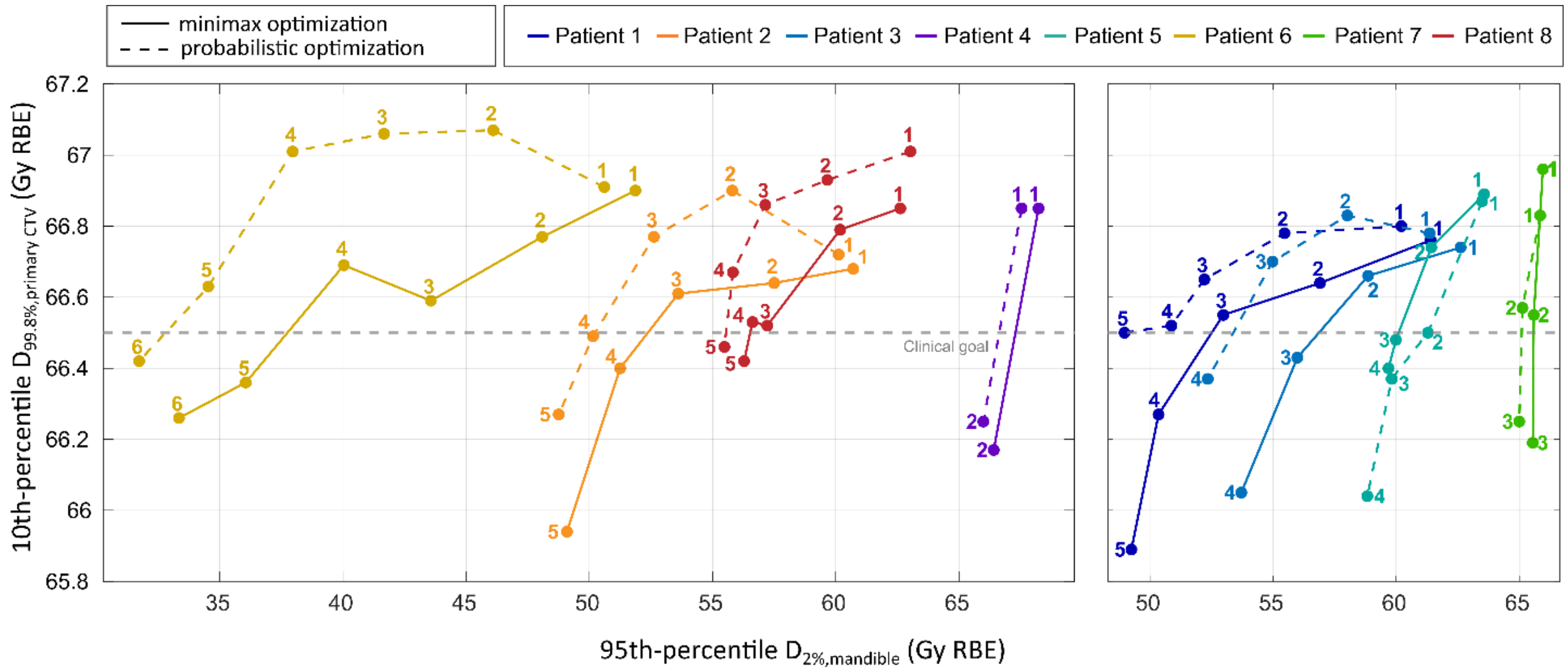

***Figure 4.*** *The full trade-off plots for all simplified oropharyngeal cancer plans. Patients are divided between two panels for readability. Trade-off curves show the relationship between 10th-percentile* $D_{99.8\%,primary\ CTV}$ *and 95th-percentile* $D_{2\%,mandible}$ *for direct probabilistic optimization and minimax optimization. Each numbered point represents a plan generated during the trade-off analysis. Plans labelled with the same number were used for the primary comparison because they were generated at the same trade-off step and had comparable primary-CTV dose homogeneity.*

*Nasopharyngeal cancer plans*

The patient-specific trade-off curves for three nasopharyngeal cancer patients are shown in Figure 5. One patient was excluded from the matched comparison because probabilistic evaluation of the original clinical plan showed that the predefined probabilistic target-coverage goal had already been achieved. The matched comparison therefore included the remaining three patients.

Plans from the same trade-off step had comparable primary-CTV dose homogeneity. In the primary matched comparison, direct probabilistic optimization showed a slightly less favorable trade-off between probabilistic target coverage and critical-OAR sparing in the two patients with an optic nerve as the critical OAR, while both optimization methods showed comparable trade-offs in the patient with the brainstem surface as the critical OAR.

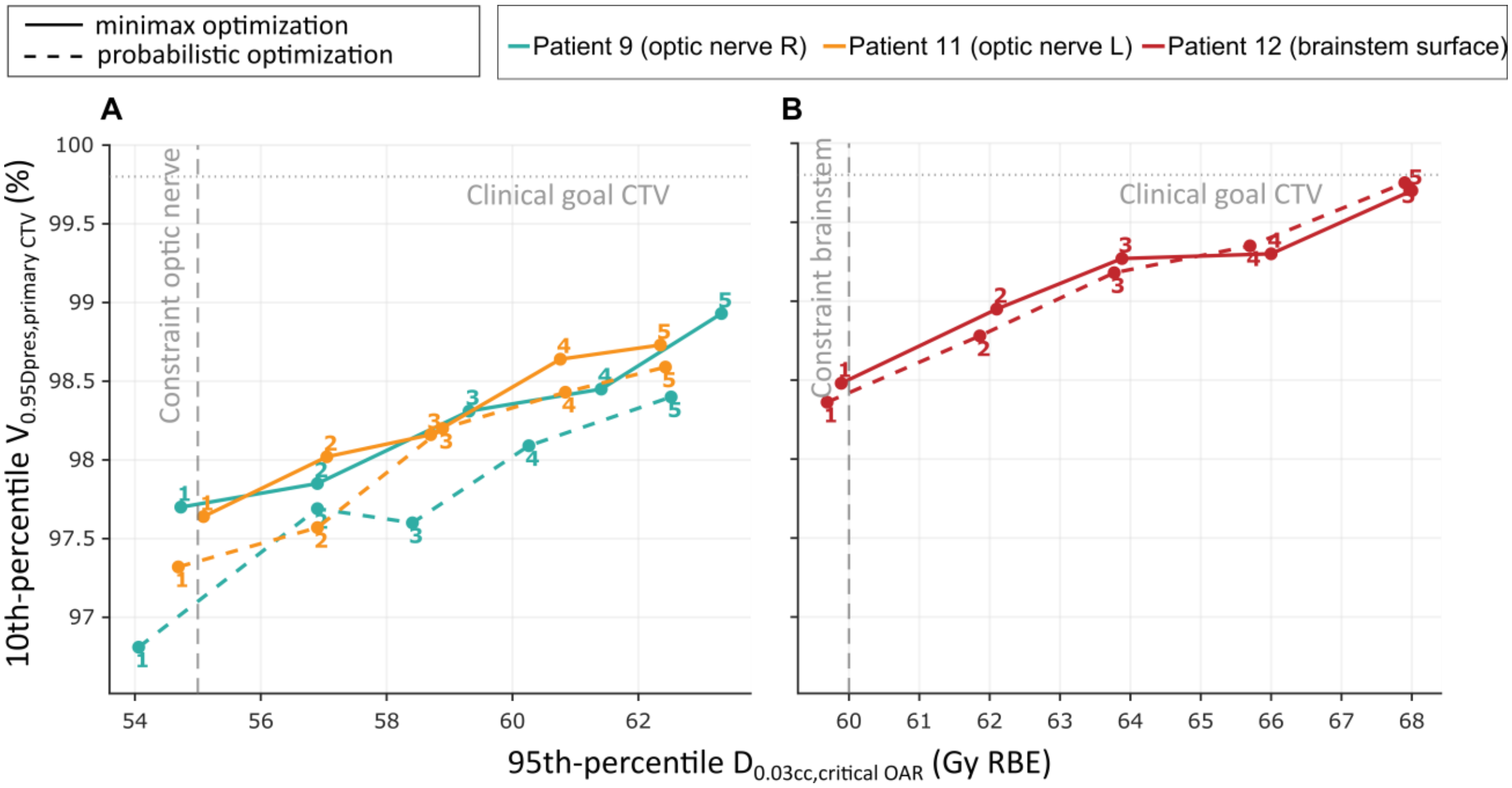

***Figure 5.*** *The full trade-off plots for three simplified nasopharyngeal cancer plans. Trade-off curves show the relationship between 10th-percentile $V_{0.95Dpres,primary\ CTV}$ and 95th-percentile $D_{0.03cc,critical\ OAR}$ for direct probabilistic optimization and minimax optimization. Each numbered point represents a plan generated during the trade-off analysis. Plans labelled with the same number were used for the primary comparison because they were generated at the same trade-off step and had comparable primary-CTV dose homogeneity.*

**Experiment 2: clinically representative plan comparison**

*Oropharyngeal cancer patients*

Before scaling, the median 10th-percentile $D_{99.8\%,primary\ CTV}$ was 66.60 Gy RBE for direct probabilistic optimization and 66.59 Gy RBE for minimax optimization. The calibrated setup robustness parameter $r_{setup}$ had a median value of 1.9 mm (range: 1.6–2.2 mm). Patient-specific values are provided in Supplementary Materials 3.

After matching the plans for probabilistic target coverage, the main comparison showed that direct probabilistic optimization reduced the 95th-percentile $D_{2\%,mandible}$ in seven of eight oropharyngeal cases. The median reduction was 0.70 Gy RBE and the median reduction in the average $D_{mean}$ across the non-critical OARs was 0.15 Gy RBE (Figure 6).

For four patients, the alternative objective configuration shifted optimization priority from mandible sparing toward reduction of the mean doses to the non-critical OARs. The advantage of direct probabilistic optimization was maintained under this strategy, with a median reduction in mean dose to the non-critical OARs of 0.28 Gy RBE compared with 0.20 Gy RBE for the original strategy (Supplementary Materials 6).

The reduction in $D_{mean}$ for the non-critical OARs varied across structures, as shown in Figure 6, with median reduction ranging from 0.07 to 0.67 Gy RBE.

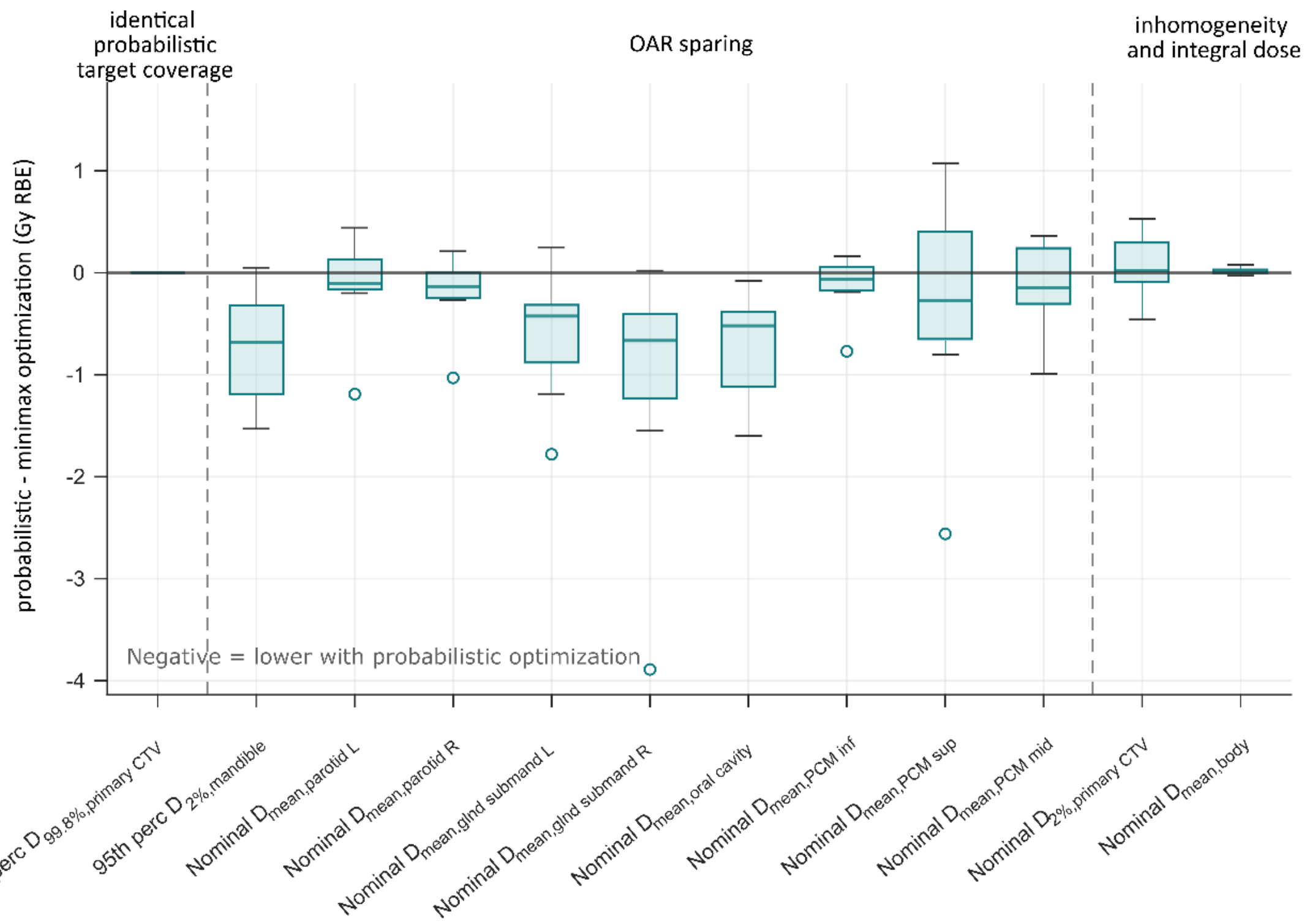

***Figure 6.*** *Results for the clinically representative oropharyngeal plans. Direct probabilistic optimization and minimax optimization were compared at matched probabilistic CTV coverage. Differences in OAR dose, target dose homogeneity and mean body dose are shown for each patient. The same results with patient-specific data points shown separately are provided in Supplementary Materials 5.*

The complete percentile distributions of $V_{0.95Dpres,\ primary\ CTV}$ are shown in Figure 7. For each percentile, values were averaged across the eight patients. Although the plans were matched at the 10th-percentile target-coverage endpoint, the distributions differed outside this matching point. Direct probabilistic optimization produced slightly lower $V_{0.95Dpres,CTV}$ values at percentiles below the 10th-percentile.

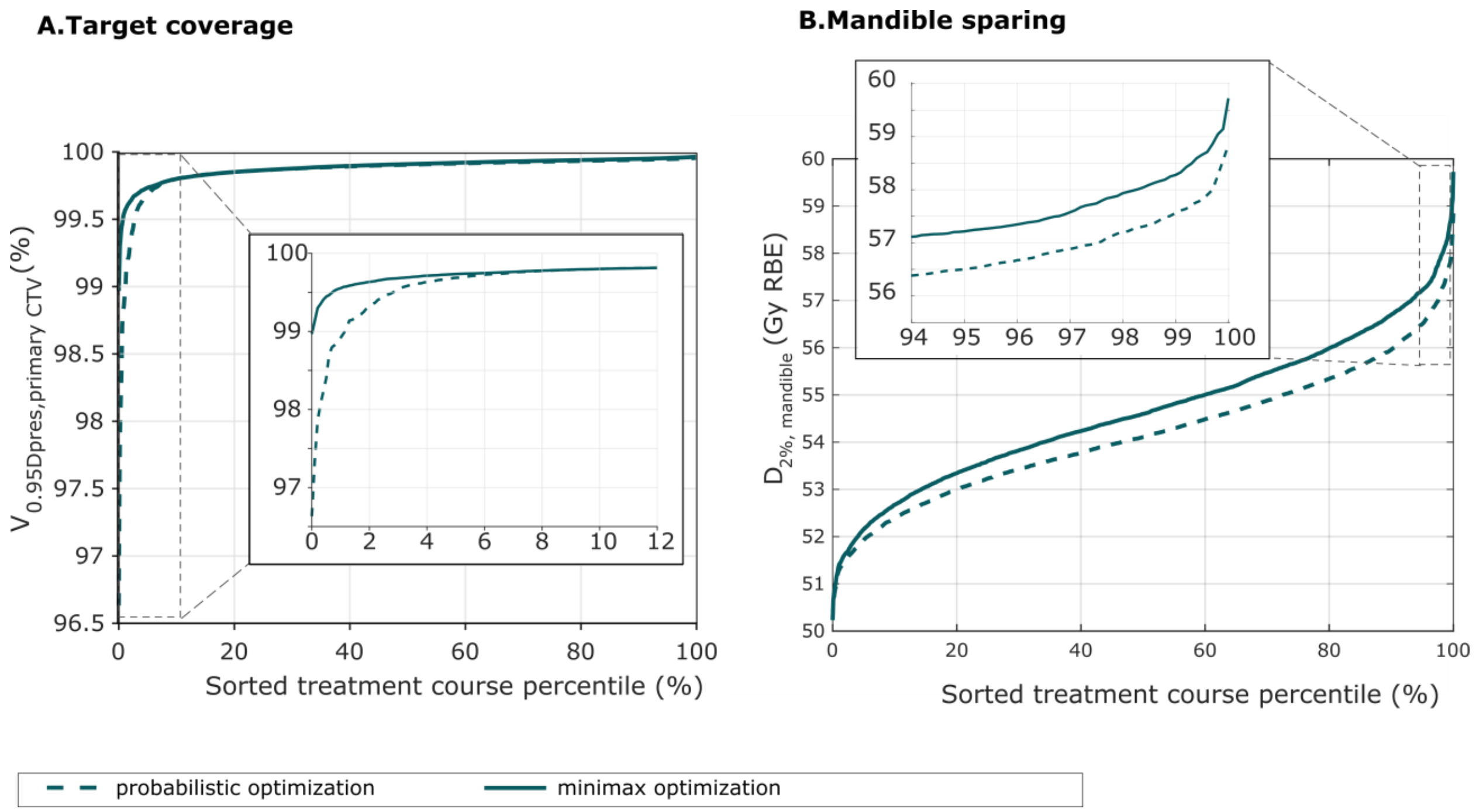

***Figure 7.*** *Full percentile plots for the clinical oropharyngeal cancer plans. The percentile distributions of $V_{0.95Dpres,primary\ CTV}$ (A) and $D_{2\%,mandible}$ (B) across simulated treatment courses are shown for direct probabilistic optimization and minimax optimization. Plans were matched at a 10th-percentile $D_{99.8\%,primary\ CTV}$ of 66.5 Gy RBE, corresponding to a 10th-percentile $V_{0.95Dpres,primary\ CTV}$ of 99.8%. Values represent the average across eight patients.*

*Nasopharyngeal cancer patients*

After direct probabilistic optimization, the achieved 95th-percentile $D_{0.03cc}$ of the patient-specific critical OAR was 54.43 Gy RBE for the right optic nerve (patient 9), 54.88 Gy RBE for the left optic nerve (patient 11) and 59.38 Gy RBE for the brainstem surface (patient 12).

For minimax optimization, the calibrated setup robustness parameters $r_{setup}$ were 2.5, 2.1 and 2.0 mm for patients 9, 11 and 12, respectively. For these plans, the 95th-percentile $D_{0.03cc}$ was 54.80 Gy RBE for the right optic nerve (patient 9), 54.86 Gy RBE for the left optic nerve (patient 11) and 59.71 Gy RBE for the brainstem surface (patient 12).

After scaling, at the matched 95th-percentile $D_{0.03cc,critical\ OAR}$, the effect of direct probabilistic optimization on probabilistic target coverage was mixed. The 10th-percentile $V_{0.95Dpres,primary\ CTV}$ was increased by 0.85, decreased by 0.79 and decreased by 0.03 percentage points for patients 9, 11 and 12, respectively (Figure 8).
The average mean doses to the other OARs are increased by 0.28, decreased by 0.20 and increased by 0.20 Gy RBE for patients 9, 11 and 12, respectively.

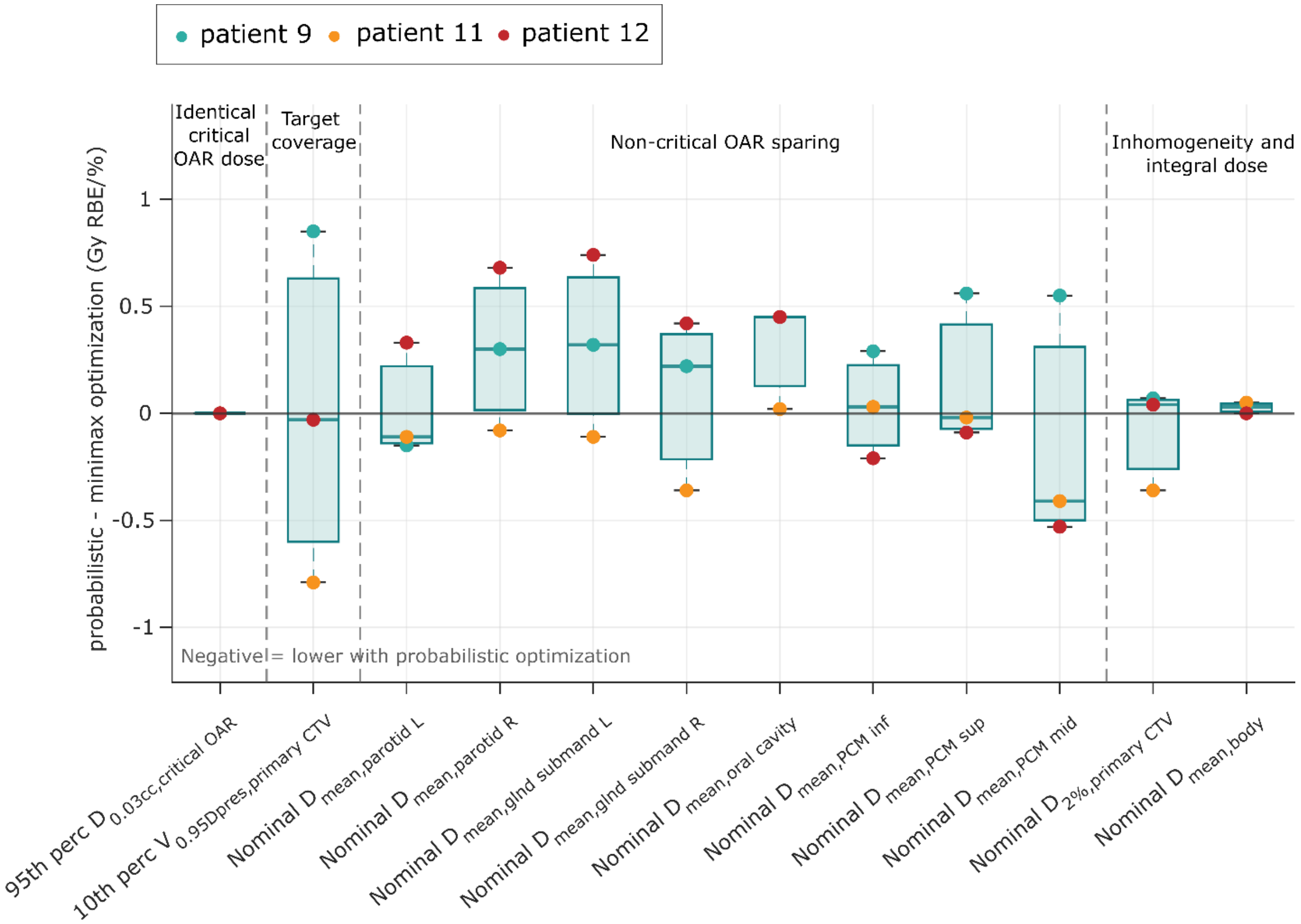

***Figure 8.*** *Results for the clinically representative nasopharyngeal plans. Direct probabilistic optimization and minimax optimization were compared at matched probabilistic critical OAR dose. Differences in target coverage, non-critical OAR dose, target dose homogeneity and mean body dose are shown for each patient.*

---

## Discussion

This study investigated whether direct probabilistic optimization can consistently steer treatment plans toward clinical goals at specified probability levels and improve the trade-off between target coverage and OAR sparing compared with probabilistically calibrated minimax optimization. In the oropharyngeal and nasopharyngeal cancer cases, direct probabilistic optimization steered treatment plans to within 0.27 and 0.62 Gy RBE, respectively, of the specified probabilistic clinical goals before final dose scaling.

In seven of eight oropharyngeal cases, direct probabilistic optimization modestly reduced OAR dose while maintaining matched probabilistic target coverage. The median reduction in the 95th-percentile $D_{2\%,\text{mandible}}$ was 0.7 Gy RBE, while the mean dose to the non-critical OARs ranged from a median reduction of 0.07 Gy RBE ($D_{\text{mean,musc constrict I}}$) to a median of 0.67 Gy RBE ($D_{\text{mean,glnd submand R}}$). In contrast, no consistent benefit for target coverage was found for the nasopharyngeal cancer cases.

The improved OAR sparing in the oropharyngeal cases may reflect the closer alignment between the optimization and evaluation criteria. While direct probabilistic optimization explicitly incorporates the specified probability levels during optimization, minimax optimization is driven by the worst-performing scenarios within a limited predefined scenario set. In addition, by accounting for fractionation, direct probabilistic optimization allows the effects of random setup errors to partially compensate across fractions in the accumulated dose distribution. Consequently, direct probabilistic optimization may allow more effective use of the available optimization freedom by reducing the influence of uncertainty scenarios that have limited relevance to the selected probabilistic endpoints.

The absence of consistent benefit in the nasopharyngeal cases may be related to the challenges of optimizing dose constraints for small critical OARs while evaluating the 95th-percentile $D_{0.03cc}$ constraints. These constraints combine a small number of voxels with an upper-tail percentile, making them particularly sensitive to numerical and sampling uncertainty. Discrepancies may arise from several sources, including the use of a discrete setup- and range-error grid during optimization, differences between dose calculation methods during optimization and evaluation and the use of an MC uncertainty of 1%. These challenges are not specific to direct probabilistic optimization and may also affect other optimization approaches. Using less extreme DVH metrics than $D_{0.03cc}$ metrics could simplify optimization, although the clinical impact of using alternatives requires further investigation. While the probability levels specified in the probabilistic optimization constraints in this study corresponded to those of the clinical goals, slightly higher probability levels may be needed during optimization to account for numerical and sampling uncertainties

Beyond dosimetric benefit, an advantage of direct probabilistic optimization is its ability to steer treatment plans directly toward predefined probabilistic clinical goals. In contrast, minimax optimization does not explicitly optimize these outcomes and therefore requires iterative, patient-specific calibration of the setup robustness parameter, as illustrated by the different calibrated values of $r_{setup}$.

Fredriksson et al. applied the direct probabilistic optimization method to a single PBS brain case and reported improved target coverage despite adhering to $D_{0.03cc}$ constraints for critical OARs [15]. A key difference is the comparison with the minimax method. The previous study used a 3.4-mm setup robustness setting, whereas in our study the minimax setting was individually calibrated using probabilistic evaluation to match the same probabilistic OAR goal, resulting in robustness settings of 2.0 to 2.5 mm for the nasopharyngeal cases. This calibration may have improved the alignment of minimax optimization with the desired probabilistic outcomes, potentially leaving less room for improvement with direct probabilistic optimization.

From a clinical perspective, direct probabilistic optimization offers the potential for more automated treatment planning by directly steering plans toward predefined probabilistic clinical goals. Direct probabilistic optimization showed dosimetric benefits for the oropharyngeal cases, where improved OAR sparing was achieved at matched 10th-percentile target coverage. However, the clinical relevance of the observed dose reductions remains uncertain, as their effect on complication probabilities or clinical outcomes was not investigated. Furthermore, the current implementation remains computationally demanding, requiring approximately eight times the optimization time of minimax optimization on a Dell workstation with an Intel Core i9-13950HX 2.2 GHz.
An additional consideration is that matching plans at a single probabilistic target-coverage endpoint does not necessarily result in equivalent target coverage across the entire probability distribution. As shown in Figure 7, direct probabilistic optimization produced slightly lower target-coverage values below the matched 10th percentile. Thus, the improvements in OAR sparing at the selected probability level were accompanied by reduced target coverage at more extreme uncertainty outcomes. Therefore, the choice of probability levels and optimization criteria should be carefully considered when defining probabilistic clinical goals.

Future research should focus on improving the balance between computational efficiency and optimization accuracy. In particular, more accurate steering toward upper-tail $D_{0.03cc}$ metrics appears important. The choice of optimization settings, including the discretization of setup and range uncertainties, may influence optimization performance and require adaptation to different clinical applications. A denser uncertainty grid could improve the representation of the tails of the uncertainty distributions, although this would increase computational burden. Alternatively, selecting setup- and range-error combinations based on a joint four-dimensional probability criterion, rather than truncating the setup-error grid independently, could reduce the number of unlikely combinations while allowing finer discretization. Other potential strategies include improved interpolation between precalculated dose distributions. Different uncertainty models, such as PCE, incorporated directly into the optimization process [21] could also be researched, although similar difficulties with extreme DVH metrics may occur. Adaptive optimization approaches, in which computational accuracy is progressively increased toward convergence, may further improve accuracy without proportionally increasing computation time.

## Conclusion

Direct probabilistic optimization enabled IMPT plans to be steered toward predefined probabilistic clinical goals and provided modest improvements in OAR sparing at matched target coverage in most oropharyngeal patients compared with probabilistically calibrated minimax optimization. No consistent advantage was observed in nasopharyngeal patients, potentially reflecting the challenges of steering on extreme dose-volume metrics for small critical OARs. Further methodological refinement, validation in larger patient cohorts and reduction of computational burden are needed before clinical implementation.

**Supplementary Materials 1**

***Table S1.*** *Comparison between DVH outcomes predicted by Polynomial chaos expansion (PCE) and Sparse probabilistic evaluation (SPE). PCE was used to determine the dose distribution under varying error scenarios. The dose* $D_i$ *to each voxel is dependent on a setup error* $\vec{\xi} = (\xi_x, \xi_y, \xi_z)$ *and a relative range error* $\rho$ *and was approximated by the analytical series expansion* $D_i(\vec{\xi}, \rho) = \sum_{k=0}^{p} a_{i,k}\, \Psi_k(\vec{\xi}, \rho)$*, with expansion coefficients* $a_{i,k}$ *and multi-dimensional Hermite polynomials* $\Psi_k(\vec{\xi}, \rho)$ *. These coefficients were determined through linear regression using 208 input-output combinations, where the input was an error scenario* $(\vec{\xi}, \rho)$ *and the output was the corresponding dose* $D_i(\vec{\xi}, \rho)$ *for each voxel* $i$ *calculated within the treatment planning system (TPS) with a MC uncertainty of 1%. Using the resulting PCE model, 10,000 fractionated treatment courses were simulated, assuming the same setup- and range-error standard deviations as used for sparse probabilistic evaluation (SPE).*

| | | Value PCE (Gy RBE/%) | Value SPE (Gy RBE) | Difference (Gy RBE/%) |
|---|---|---|---|---|
| **Patient 9** | **10th perc $D_{99.8\%,CTV}$** | 50.72 | 50.72 | 0.00 |
| | **10th perc $V_{0.95Dpres}$** | 94.73 | 94.83 | 0.10 |
| | **95th perc $D_{0.03cc,brainstem\ Core}$** | 52.27 | 51.99 | -0.28 |
| | **95th perc $D_{0.03cc,brainstem\ Surf}$** | 60.09 | 59.82 | -0.27 |
| | **95th perc $D_{0.03cc,Optic\ chiasm}$** | 53.23 | 53.09 | -0.14 |
| | **95th perc $D_{0.03cc,optic\ nrv\ L}$** | 51.29 | 51.38 | 0.09 |
| | **95th perc $D_{0.03cc,optic\ nrv\ R}$** | 55.23 | 55.11 | -0.12 |
| **Patient 11** | **10th perc $D_{99.8\%,CTV}$** | 49.94 | 50.00 | 0.06 |
| | **10th perc $V_{0.95Dpres}$** | 94.39 | 94.57 | 0.18 |
| | **95th perc $D_{0.03cc,brainstem\ Core}$** | 51.64 | 51.50 | -0.14 |
| | **95th perc $D_{0.03cc,brainstem\ Surf}$** | 58.89 | 58.87 | -0.02 |
| | **95th perc $D_{0.03cc,Optic\ chiasm}$** | 54.09 | 54.02 | -0.07 |
| | **95th perc $D_{0.03cc,optic\ nrv\ L}$** | 55.05 | 55.05 | 0.00 |
| | **95th perc $D_{0.03cc,optic\ nrv\ R}$** | 47.76 | 47.49 | -0.27 |
| **Patient 12** | **10th perc D99.8%,CTV** | 59.63 | 59.48 | -0.15 |
| | **10th perc V0.95Dpres** | 98.05 | 98.17 | 0.12 |
| | **95th perc D0.03cc,brainstem Core** | 52.50 | 52.28 | -0.22 |
| | **95th perc D0.03cc,brainstem Surf** | 60.34 | 60.32 | -0.02 |
| | **95th perc D0.03cc,Optic chiasm** | 53.69 | 53.68 | -0.01 |
| | **95th perc D0.03cc,optic nrv L** | 52.00 | 52.20 | 0.2 |

| | | | |
|---|---|---|---|
| **95th perc D0.03cc,optic nrv R** | 53.23 | 53.16 | -0.07 |

**Supplementary Materials 2**

***Table S2.*** *Optimization objectives and constraints for an example oropharyngeal cancer patient in Experiment 1. The simplified optimization problem included probabilistic target-coverage constraints for the primary and nodal CTVs and a probabilistic D2%,mandible objective, together with the non-probabilistic auxiliary objectives used for dose shaping. The same objective and constraint list was used for direct probabilistic optimization and probabilistically calibrated minimax optimization.*

| Constraint | ROI | Description | Robust / probabilistic | Weight |
|---|---|---|---|---|
| — | opt_Ext | Dose fall-off [H] 71.00 Gy (RBE); [L] 10.00 Gy (RBE); low-dose distance 2.00 cm | — | 1 |
| — | opt_Ext | Max dose 282.00 Gy (RBE); all beams | — | 10000 |
| — | opt_Ext | Max dose 74.90 Gy (RBE) | — | 100 |
| — | opt_CTV_R | Min dose 0.00 Gy (RBE); beam 'B4_G310C0RS3' | — | 0 |
| — | opt_CTV_R | Min dose 0.00 Gy (RBE); beam 'B5_G260C20RS3' | — | 0 |
| — | opt_CTV_R | Min dose 0.00 Gy (RBE); beam 'B6_G200C0' | — | 0 |
| — | opt_CTV_L | Min dose 0.00 Gy (RBE); beam 'B1_G160C0' | — | 0 |
| — | opt_CTV_L | Min dose 0.00 Gy (RBE); beam 'B2_G100C340RS3' | — | 0 |
| — | opt_CTV_L | Min dose 0.00 Gy (RBE); beam 'B3_G50C0RS3' | — | 0 |
| — | opt_CTV_5425-CTV_7000+3mm | Max dose 54.25 Gy (RBE) | — | 5 |
| — | opt_CTV_5425-CTV_7000+2cm | Max dose 54.25 Gy (RBE) | — | 80 |
| — | CTV_7000 | Max dose 70.00 Gy (RBE) | — | 500 |
| — | Bone_Mandible | Max DVH 60.00 Gy (RBE) to 2.00% volume | Yes | 1000 |
| Yes | CTV_7000 | Min DVH 67.00 Gy (RBE) to 99.80% volume | Yes | — |
| Yes | CTV_5425 | Min DVH 52.00 Gy (RBE) to 99.80% volume | Yes | — |

***Table S3.*** *Optimization objectives and constraints for an example nasopharyngeal cancer patient in Experiment 1. The simplified optimization problem included probabilistic target-coverage objectives for the primary and nodal CTVs and a probabilistic constraint for the patient-specific critical OAR, together with the non-probabilistic auxiliary objectives used for dose shaping. The same objective and constraint list was used for direct probabilistic optimization and probabilistically calibrated minimax optimization.*

| **Constraint** | **ROI** | **Description** | **Robust / probabilistic** | **Weight** |
|---|---|---|---|---|
| — | opt_Ext | Max dose 282.00 Gy (RBE), All beams | — | 10000 |
| — | opt_Ext | Max dose 74.50 Gy (RBE) | — | 100 |
| — | opt_Ext | Dose fall-off [H] 71.00 Gy (RBE), [L] 0.00 Gy (RBE); low-dose distance 3.00 cm | — | 5 |
| — | opt_CTV_R | Min dose 0.00 Gy (RBE), Beam 'B4_G310C0RS3' | — | 0 |
| — | opt_CTV_R | Min dose 0.00 Gy (RBE), Beam 'B5_G260C20RS3' | — | 0 |
| — | opt_CTV_R | Min dose 0.00 Gy (RBE), Beam 'B6_G200C0' | — | 0 |
| — | opt_CTV_L | Min dose 0.00 Gy (RBE), Beam 'B1_G160C0' | — | 0 |
| — | opt_CTV_L | Min dose 0.00 Gy (RBE), Beam 'B2_G100C340RS3' | — | 0 |
| — | opt_CTV_L | Min dose 0.00 Gy (RBE), Beam 'B3_G50C0RS3' | — | 0 |
| — | opt_CTV_5425-CTV_7000+3mm | Max dose 54.25 Gy (RBE) | — | 30 |
| — | opt_CTV_5425-CTV_7000+2cm | Max dose 54.25 Gy (RBE) | — | 80 |
| yes | CTV_7000 | Max dose 74.00 Gy (RBE) | — | |
| — | CTV_7000 | Min DVH 67.00 Gy (RBE) to 99.80% volume | Yes | 10000 |
| — | CTV_5425 | Min DVH 52.00 Gy (RBE) to 99.80% volume | Yes | 1000 |
| Yes | OpticNrv_R | Max dose 55.00 Gy (RBE) | Yes | — |

***Table S4.*** *Optimization objectives and constraints for an example oropharyngeal cancer patient in Experiment 2. The clinically representative optimization problem included probabilistic target-coverage constraints, a probabilistic mandible objective, objectives for the remaining OARs, and auxiliary objectives for dose shaping and target-dose control. The same objective and constraint list was used for direct probabilistic optimization and probabilistically calibrated minimax optimization.*

| Constraint | ROI | Description | Robust / probabilistic | Weight |
|---|---|---|---|---|
| — | opt_CTV_5425-CTV_7000+3mm | Max dose 54.25 Gy (RBE) | — | 5 |
| — | CTV_7000 | Max dose 70.00 Gy (RBE) | — | 500 |
| — | opt_Ext | Dose fall-off [H] 71.00 Gy (RBE), [L] 10.00 Gy (RBE); low dose distance 2.00 cm | — | 1 |
| — | opt_Ext | Max dose 282.00 Gy (RBE), all beams | — | 10000 |
| — | opt_CTV_5425-CTV_7000+2cm | Max dose 54.25 Gy (RBE) | — | 80 |
| — | opt_Ext | Max dose 74.90 Gy (RBE) | — | 100 |
| — | opt_CTV_L | Min dose 0.00 Gy (RBE), beam 'B1_G160C0' | — | 0 |
| — | opt_CTV_L | Min dose 0.00 Gy (RBE), beam 'B2_G100C340RS3' | — | 0 |
| — | opt_CTV_L | Min dose 0.00 Gy (RBE), beam 'B3_G50C0RS3' | — | 0 |
| — | opt_CTV_R | Min dose 0.00 Gy (RBE), beam 'B4_G310C0RS3' | — | 0 |
| — | opt_CTV_R | Min dose 0.00 Gy (RBE), beam 'B5_G260C20RS3' | — | 0 |
| — | opt_CTV_R | Min dose 0.00 Gy (RBE), beam 'B6_G200C0' | — | 0 |
| — | Glnd_Submand_L | Max EUD 43.54 Gy (RBE), parameter A 1 | — | 1 |
| — | Parotid_R-CTV03 | Max EUD 14.25 Gy (RBE), parameter A 1 | — | 1 |
| — | Parotid_L-CTV03 | Max EUD 14.35 Gy (RBE), parameter A 1 | — | 1 |
| — | Glnd_Submand_R-CTV03 | Max EUD 25.89 Gy (RBE), parameter A 1 | — | 1 |
| — | Oral_cavity-CTV03 | Max EUD 7.37 Gy (RBE), parameter A 1 | — | 1 |
| — | Musc_Constrict_S-CTV03 | Max EUD 29.71 Gy (RBE), parameter A 1 | — | 1 |
| — | Musc_Constrict_M-CTV03 | Max EUD 18.49 Gy (RBE), parameter A 1 | — | 1 |
| — | Musc_Constrict_I-CTV03 | Max EUD 12.24 Gy (RBE), parameter A 1 | — | 1 |
| — | Bone_Mandible | Max EUD 15.13 Gy (RBE), parameter A 1 | — | 1 |
| — | Bone_Mandible | Max DVH 50.00 Gy (RBE) to 2.00% volume | Yes | 1000 |
| — | External | Max EUD 6.88 Gy (RBE), parameter A 1 | — | 1000 |
| Yes | CTV_7000 | Min DVH 67.00 Gy (RBE) to 99.80% volume | Yes | |
| Yes | CTV_5425 | Min DVH 52.00 Gy (RBE) to 99.80% volume | Yes | |

***Table S5.*** *Optimization objectives and constraints for an example nasopharyngeal cancer patient in Experiment 2. The clinically representative optimization problem included probabilistic target-coverage objectives, probabilistic constraints for the brainstem surface, brainstem core, optic chiasm and optic nerves, objectives for the remaining OARs, and auxiliary objectives for dose shaping and target-dose control. The same objective and constraint list was used for direct probabilistic optimization and probabilistically calibrated minimax optimization.*

| Constraint | ROI | Description | Robust / probabilistic | Weight |
|---|---|---|---|---|
| — | opt_CTV_5425-CTV_7000+3mm | Max dose 54.25 Gy (RBE) | — | 30 |
| — | opt_Ext | Dose fall-off [H] 71.00 Gy (RBE), [L] 0.00 Gy (RBE), low-dose distance 3.00 cm | — | 5 |
| — | opt_Ext | Max dose 282.00 Gy (RBE), all beams | — | 10000 |
| — | opt_CTV_5425-CTV_7000+2cm | Max dose 54.25 Gy (RBE) | — | 80 |
| — | opt_Ext | Max dose 74.50 Gy (RBE) | — | 100 |
| — | opt_CTV_L | Min dose 0.00 Gy (RBE), beam 'B1_G160C0' | — | 0 |
| — | opt_CTV_L | Min dose 0.00 Gy (RBE), beam 'B2_G100C340RS3' | — | 0 |
| — | opt_CTV_L | Min dose 0.00 Gy (RBE), beam 'B3_G50C0RS3' | — | 0 |
| — | opt_CTV_R | Min dose 0.00 Gy (RBE), beam 'B4_G310C0RS3' | — | 0 |
| — | opt_CTV_R | Min dose 0.00 Gy (RBE), beam 'B5_G260C20RS3' | — | 0 |
| — | opt_CTV_R | Min dose 0.00 Gy (RBE), beam 'B6_G200C0' | — | 0 |
| — | Parotid_R-CTV03 | Max EUD 11.50 Gy (RBE), parameter A 1 | — | 1 |
| — | Parotid_L-CTV03 | Max EUD 8.20 Gy (RBE), parameter A 1 | — | 1 |
| — | Glnd_Submand_R-CTV03 | Max EUD 21.00 Gy (RBE), parameter A 1 | — | 1 |
| — | Glnd_Submand_L-CTV03 | Max EUD 27.00 Gy (RBE), parameter A 1 | — | 1 |
| — | Oral_cavity-CTV03 | Max EUD 12.00 Gy (RBE), parameter A 1 | — | 1 |
| — | Musc_Constrict_S-CTV03 | Max EUD 40.00 Gy (RBE), parameter A 1 | — | 1 |
| — | Musc_Constrict_M-CTV03 | Max EUD 27.00 Gy (RBE), parameter A 1 | — | 1 |
| — | Musc_Constrict_I-CTV03 | Max EUD 23.50 Gy (RBE), parameter A 1 | — | 1 |
| yes | CTV_7000 | Max dose 74.00 Gy (RBE) | — | |
| — | CTV_7000 | Min DVH 67.00 Gy (RBE) to 99.80% volume | Yes | 10000 |
| — | CTV_5425 | Min DVH 52.00 Gy (RBE) to 99.80% volume | Yes | 1000 |
| — | External | Max EUD 5.61 Gy (RBE), parameter A 1 | — | 1000 |
| Yes | Brainstem_Surf | Max dose 60.00 Gy (RBE) | Yes | |
| Yes | Brainstem_Core | Max dose 54.00 Gy (RBE) | Yes | |
| Yes | OpticChiasm | Max dose 55.00 Gy (RBE) | Yes | |
| Yes | OpticNrv_L | Max dose 55.00 Gy (RBE) | Yes | |
| Yes | OpticNrv_R | Max dose 55.00 Gy (RBE) | Yes | |
| Yes | Bone_Mandible | Max DVH 70.00 Gy (RBE) to 2.00% volume | Yes | |

**Supplementary Materials 3**

***Table S6.*** *Patient-specific 10th-percentile $D_{99.8\%,\ primary\ CTV}$ values obtained with direct probabilistic optimization and probabilistically calibrated minimax optimization for the eight oropharyngeal cancer patients in Experiments 1 and 2, together with the corresponding calibrated setup robustness parameters $r_{setup}$. Values are reported before dose scaling.*

| | **Experiment 1** | | | **Experiment 2** | | |
|---|---|---|---|---|---|---|
| **Patient** | **Probabilistic $D_{99.8\%,CTV}$ (Gy RBE)** | **Minimax $D_{99.8\%,CTV}$ (Gy RBE)** | **Robustness setting (mm)** | **Probabilistic $D_{99.8\%,CTV}$ (Gy RBE)** | **Minimax $D_{99.8\%,CTV}$ (Gy RBE)** | **Robustness setting (mm)** |
| 1 | 66.8 | 66.76 | 1.5 | 66.56 | 66.53 | 1.6 |
| 2 | 66.72 | 66.68 | 1.5 | 66.58 | 66.63 | 1.6 |
| 3 | 66.78 | 66.74 | 1.7 | 66.64 | 66.53 | 2.2 |
| 4 | 66.85 | 66.85 | 1.8 | 66.7 | 66.63 | 2 |
| 5 | 66.87 | 66.89 | 1.8 | 66.56 | 66.62 | 1.8 |
| 6 | 66.91 | 66.9 | 1.5 | 66.57 | 66.56 | 1.8 |
| 7 | 66.83 | 66.96 | 2.0 | 66.77 | 66.75 | 2 |
| 8 | 67.01 | 66.85 | 2.0 | 66.62 | 66.54 | 2 |
| Median | 66.84 | 66.85 | 1.8 | 66.6 | 66.59 | 1.9 |
| Min | 66.72 | 66.68 | 1.5 | 66.56 | 66.53 | 1.6 |
| Max | 67.01 | 66.96 | 2.0 | 66.77 | 66.75 | 2.2 |

## Supplementary Materials 4

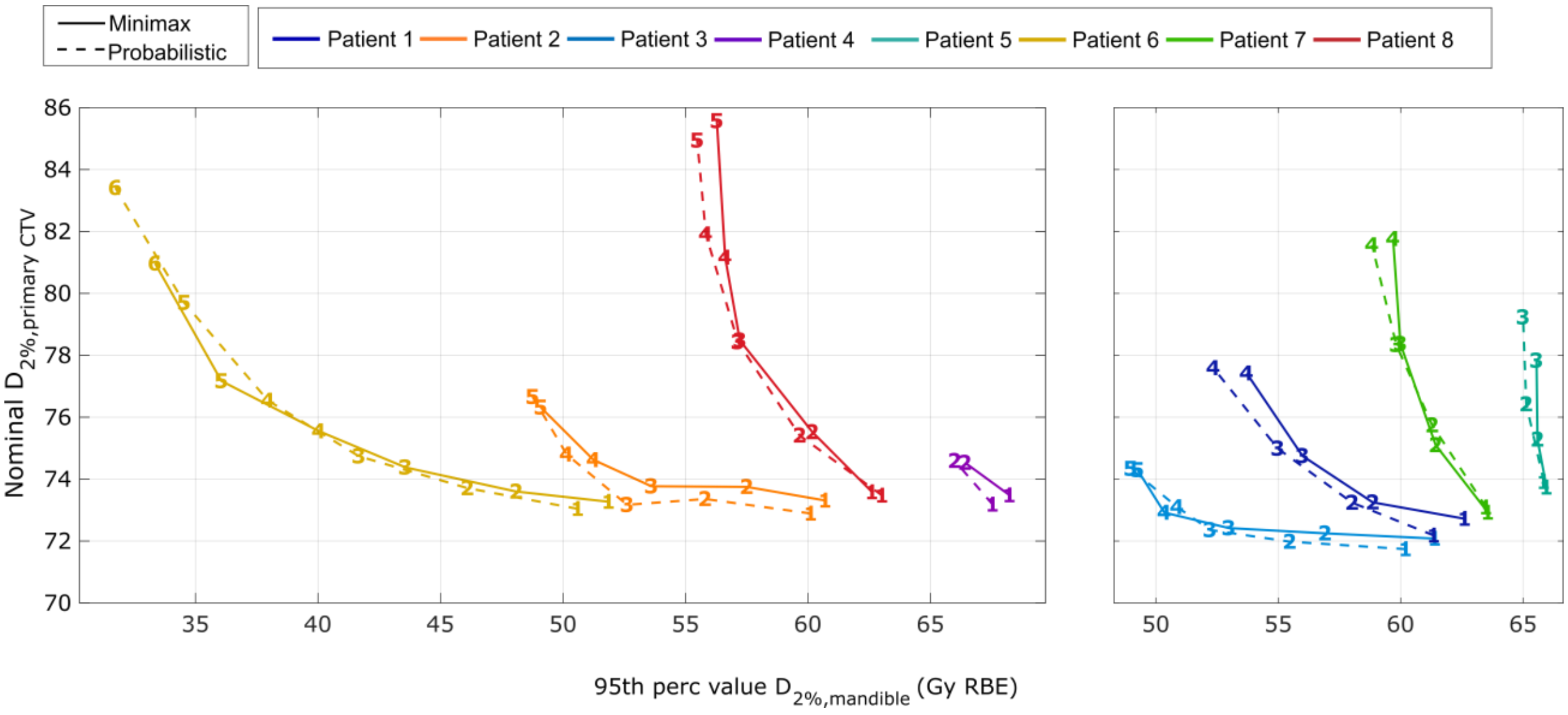

***Figure S1.*** *Primary-CTV dose inhomogeneity across the simplified oropharyngeal trade-off analysis. Nominal* $D_{2\%,primary\ CTV}$ *is plotted against the 95th-percentile* $D_{2\%,mandible}$ *for direct probabilistic optimization and minimax optimization in all eight oropharyngeal cancer patients. Each numbered point represents a successive step along the trade-off curve. Plans labelled with the same number correspond to the same trade-off step used for the matched comparisons in Figure 4.*

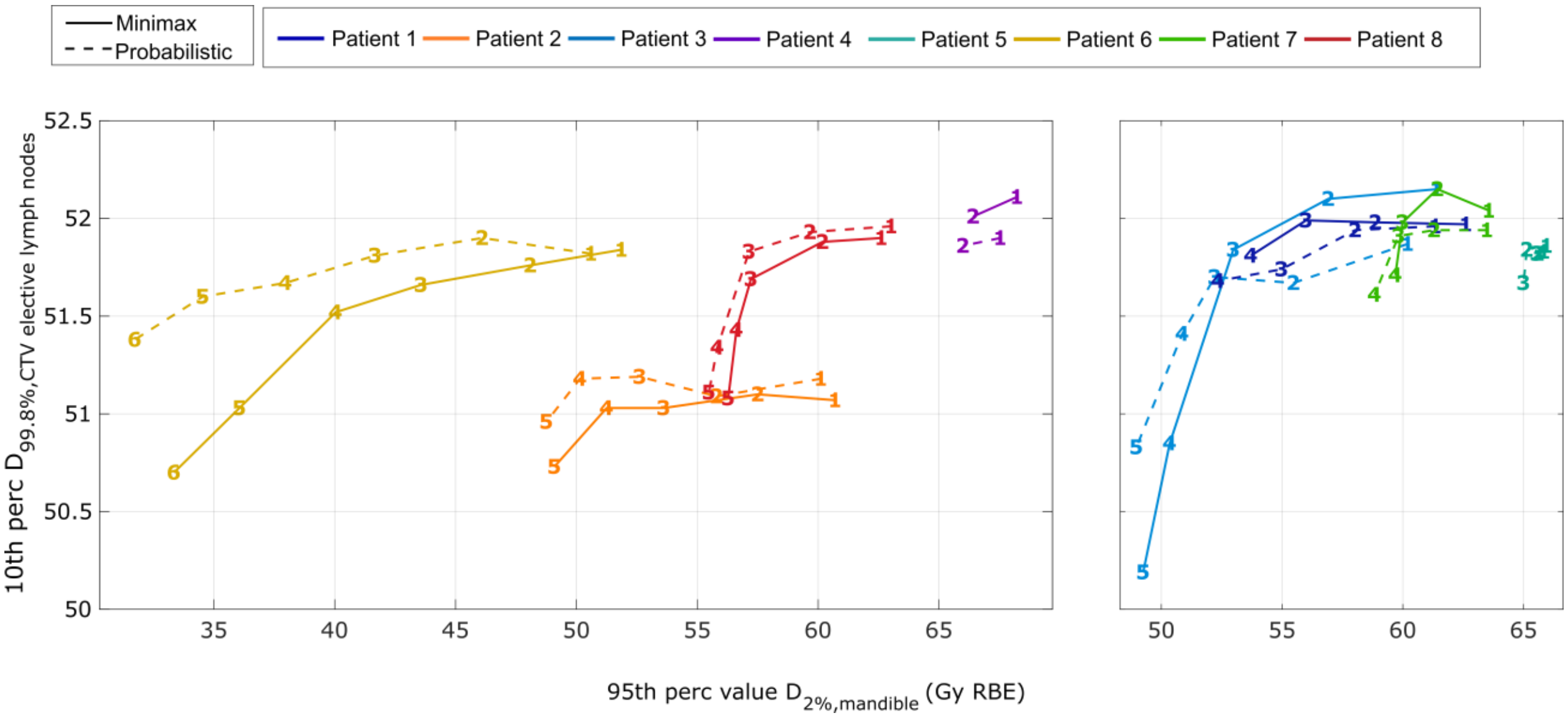

***Figure S2.*** *Nodal-CTV target coverage across the simplified oropharyngeal trade-off analysis. The 10th-percentile* $D_{99.8\%,nodal\ CTV}$ *is plotted against the 95th-percentile* $D_{2\%,mandible}$ *for direct probabilistic optimization and minimax optimization in all eight oropharyngeal cancer patients. Each numbered point represents a successive step along the trade-off curve, with equally numbered plans corresponding to the same trade-off step.*

**Supplementary Materials 5**

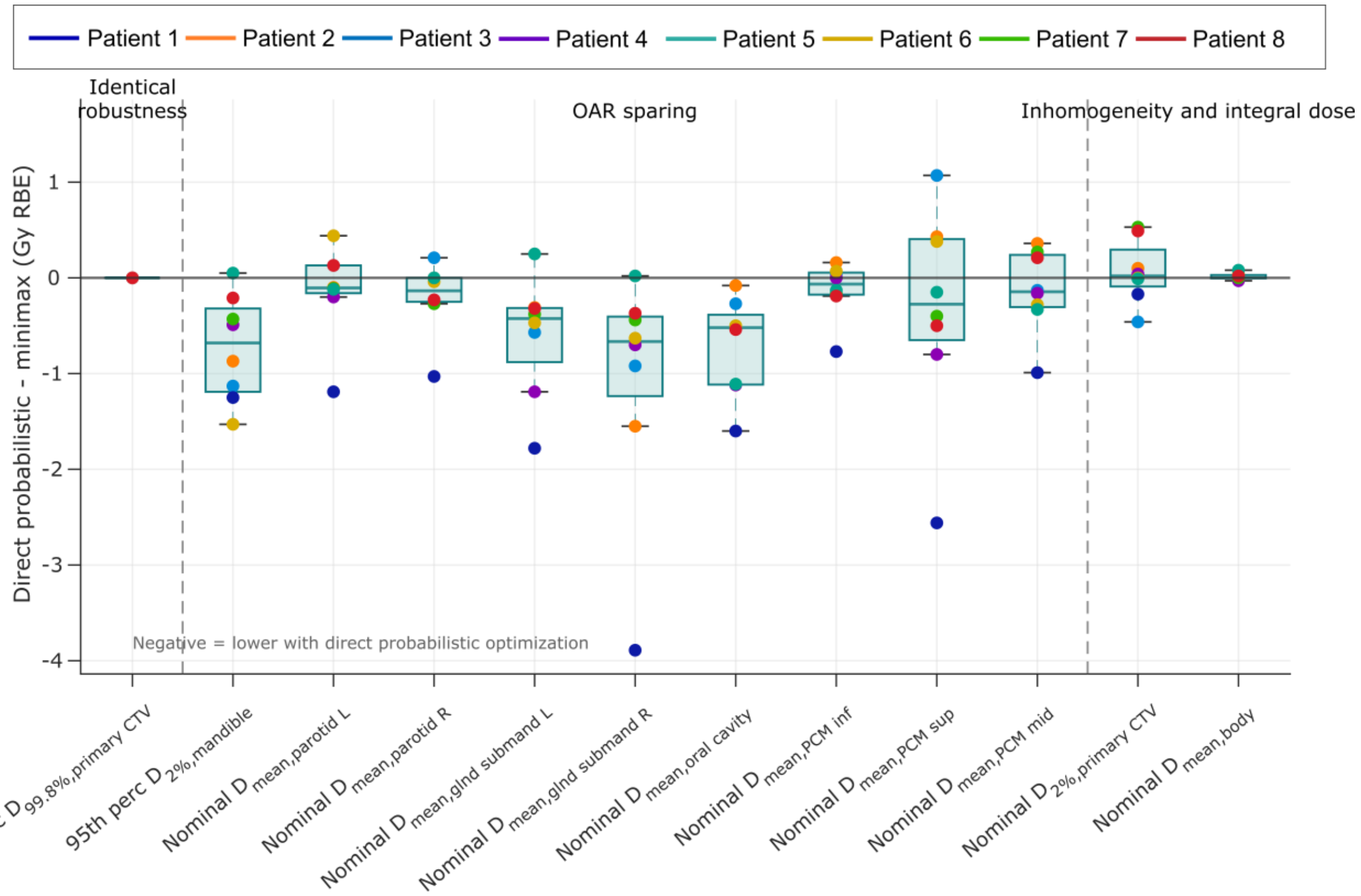

***Figure S3.*** *Patient-specific results for the clinically representative oropharyngeal cancer plans. The same results as in Figure 6 are shown, with individual patient data points overlaid on the boxplots. Differences between direct probabilistic optimization and minimax optimization are shown after matching plans for probabilistic primary-CTV coverage. Negative values indicate lower dose with direct probabilistic optimization. Individual colored markers represent the eight patients, and boxplots summarize the patient distribution. Results are shown for the 95th-percentile* $D_{2\%,mandible}$*, nominal mean doses to the non-critical OARs, nominal* $D_{2\%,primary\ CTV}$ *and nominal mean body dose.*

**Supplementary Materials 6**

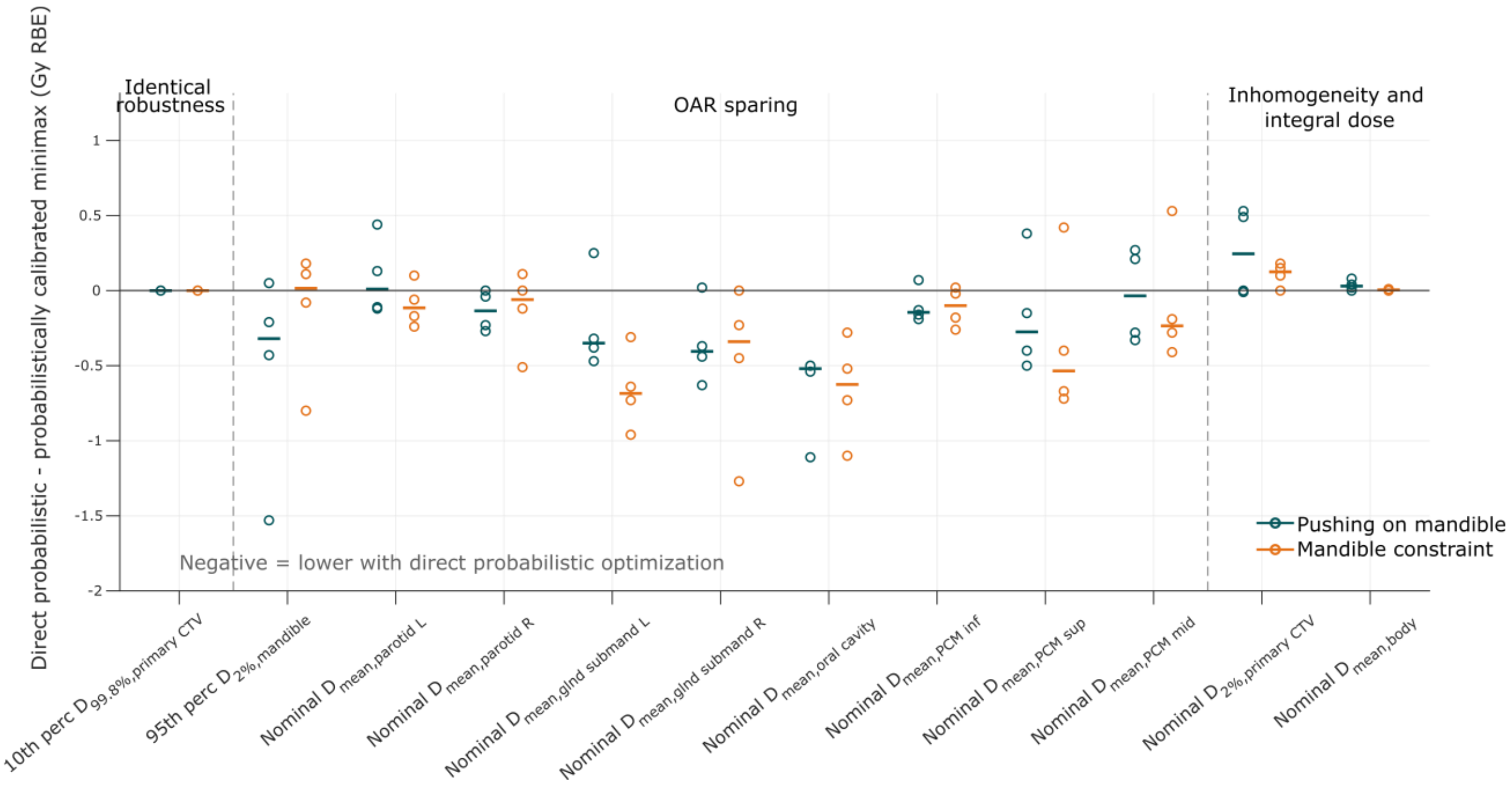

***Figure S4.*** *Comparison of the two optimization strategies for the four oropharyngeal cancer patients. Differences between direct probabilistic optimization and minimax optimization are shown after matching plans for probabilistic primary-CTV coverage. The original strategy prioritized mandible sparing, whereas the alternative strategy applied an upper* $D_{2\%,mandible}$ *constraint and used the remaining optimization freedom to reduce mean doses to the non-critical OARs. Negative values indicate lower dose with direct probabilistic optimization. Horizontal bars indicate the average difference across the four patients for each strategy.*

**Supplementary Materials 7**

The remaining $\alpha$% of treatment courses were used to calculate the aggregated function value using a generalized mean:

$$M_p(x_1, \dots, x_n) = \left(\frac{1}{n}\sum_{i=1}^{n} x_i^p\right)^{1/p}$$

For the probabilistic constraints we use p=1. The aggregated constraint violation should approach zero:

$$M_1(x_1, \dots, x_n) = \frac{1}{n}\sum_{i=1}^{n} x_i \to 0$$

For the probabilistic objectives we use p=8. The aggregated objective-function value is minimized:

$$minimize\ M_8(x_1, \dots, x_n) = \left(\frac{1}{n}\sum_{i=1}^{n} x_i^8\right)^{1/8}$$

Here, $x_i$ is the function value for retained treatment course $i$, and $n$ is the number of retained treatment courses (e.g., 900 for a 90% probability level and 1,000 simulated treatment courses).